\documentclass[a4paper,english,prb, twocolumn]{revtex4}
\usepackage{ae} 
\usepackage[T1]{fontenc}
\usepackage[ansinew]{inputenc}
\usepackage{amsmath}
\usepackage{amssymb}
\usepackage{amsfonts}
\usepackage{natbib}
\usepackage[dvips]{graphicx}
\usepackage{graphics}
\usepackage{eurosym}
\usepackage{babel}
\usepackage{multirow}
\usepackage{color}

\begin{document}

\newcommand{\comm}[1]{}
\newcommand{\ffbox}[1]{}
\newcommand{\oldtext}[1]{}
\newcommand{\comment}[1]{}

\title{Kohn--Sham potential with discontinuity for band gap materials}
\author{M.~Kuisma$^a$, J.~Ojanen$^a$, J.~Enkovaara$^b$ and T.T.~Rantala$^a$}
\affiliation{$^a$Department of Physics, Tampere University of Technology,
P.O. Box 692, FI-33101 Tampere, Finland\\
$^b$ CSC -- IT Center for Science Ltd.
P.O. Box 405
FI-02101 Espoo, Finland}

\date{\today}

\begin{abstract}
We model a Kohn--Sham potential with a discontinuity at integer particle numbers derived from the GLLB approximation of Gritsenko {\it et al.}
We evaluate the Kohn--Sham gap and the discontinuity to obtain the quasiparticle gap. 
This allows us to compare the Kohn--Sham gaps to those obtained by accurate many-body perturbation theory based optimized potential methods.
In addition, the resulting quasiparticle band gap is compared to experimental gaps.
In the GLLB model potential, the exchange--correlation hole is modeled using a GGA energy 
density and the response of the hole to density variations is evaluated by using the common-denominator 
approximation and homogeneous electron gas based assumptions. 
In our modification, we have chosen the PBEsol potential as the GGA to model the exchange hole, and add a consistent correlation potential.
The method is implemented in the GPAW code, which allows efficient parallelization to study large systems.
A fair agreement for Kohn--Sham and the quasiparticle band gaps with semiconductors and other band gap materials is obtained
with a potential which is as fast as GGA to calculate.
\end{abstract}

\maketitle

\newcommand{\rr}{{\bf r}}
\newcommand{\Ref}[1]{(\ref{#1})}
\newcommand{\ket}[1]{|#1\rangle}
\newcommand{\bra}[1]{\langle#1|}
\newcommand{\trueks}{``True'' KS}
\newcommand{\GLLBSC}{GLLB-SC}
\newcommand{\etal}{{\it et al.}}

\section{Introduction}
The Kohn--Sham density functional theory (KS-DFT) \cite{Hohenberg1964,KohnSham}
with local and semi local density approximations (LDA and GGA) has
proven to be successful in predicting total energy related properties
of many electron systems, such as crystal structures, molecular geometries and
cohesion energies.
Therefore, these simple approximations can be used to predict many of the
the ground state features of metals, semiconductors and dielectrics.

Although there is no direct physical interpretation of the KS-eigenvalues \cite{KSEnergies} 
the eigenvalue differences can be considered as zeroth order approximations to the
excitation energies \cite{PhysRevA.54.3912} or the eigenvalues itself as vertical ionization potentials.\cite{chong:1760}
In some cases, the shape of the valence and conduction 
bands also resembles the experimentally measured
ones, except for the band gap.
The physical quasiparticle gap contains, in addition to the KS band gap, the integer derivative discontinuity
of the exchange--correlation (XC) functional.\cite{ShamSchluter, KSEnergies}
As this contribution is positive and not small, the KS band gap underestimates severely the observed ones
also for potentials believed to be close to the exact KS-DFT.\cite{gruning, PhysRevB.36.6497,  PhysRevLett.56.2415}

The conventional solution to this sc.~band gap problem has been
an empirical shift, often called as "scissor operation", to correct the too small band gap
\cite{PhysRevLett.63.1719}.
Tran \etal~published a different approach with semilocal model potential 
to evaluate band gaps of solids \cite{tran:226401} by fitting 
the potential, defined with parameters, to increase KS band gap
to reproduce the experimental one. 
However, due to the Hohenberg--Kohn theorem \cite{Hohenberg1964}, 
there is only one KS
potential which yields the correct density, and it has been shown
that the accurate many body perturbation theory based KS potential yields a KS gap,
which in most cases is only little more than half of the experimental
one. \cite{ShamSchluter, PhysRevB.36.6497, gruning} 
The previous statements reflect our point of view, for we will consider these potentials
as best available references and will compare our results accordingly.

The potential discontinuity at integer particle number is only an artifact of multiplicative
KS potential and the proper quasiparticle
picture such as non-local Hartree--Fock or non-local and energy dependent GW\cite{Hedin} 
directly yields the quasiparticle band gap as a one electron energy
difference of highest occupied and lowest unoccupied levels. 
Thus, there are different approaches to obtain
good quasiparticle band gaps, KS-DFT with a multiplicative potential and others which employ non-locality, either spatial or temporal (energy dependence). 
The evaluation of the discontuity for a multiplicative
KS-potential has been stated necessary and cumbersome \cite{PhysRevB.74.161103}, but in case of GLLB it is trivial.

A general formalism to obtain the KS-potential for a given (not explicitly density dependent) energy
functional is the optimized effective potential method (OEP)\cite{Talman,
PhysRev.90.317}, where the total energy functional is minimized
with respect to variations in a multiplicative XC potential. Correlation
contributions can also be included\cite{Grabo1995141,PhysRevA.51.2005,0953-8984-10-41-007}, 
but a common approach is
sc.~exact-exchange OEP (EXX-OEP) formalism \cite{GorlingDxc}, where the Hartree--Fock exchange energy
functional is adopted as the first order term of the adiabatic perturbation theory. \cite{PhysRevA.50.196}

Several approximations have been suggested for solving the complicated
and computationally demanding OEP equations,
such as common denominator approximation based on sc.~KLI\cite{KLI} and
LHF\cite{sala:5718} potentials.
The practical calculations for large systems, however, call for more robust approaches.

In this study we present one such alternative.
We start with the computationally attractive
GLLB potential\cite{GLLB} by Gritsenko \etal,
which is a further approximation to KLI potential.
The model includes the useful properties of the electron gas as well as a discontinuity on integer particle number.
The GLLB-potential is further modified by replacing the used energy density functional to another,
more suitable for solids and adding correlation. This results in a potential which we call
\GLLBSC~(solid, correlation).

We have implemented our approach along with the GLLB
within the projector augmented wave (PAW) method in the 
real-space grid based GPAW code.\cite{GPAW}.
As a test set we consider the elemental semiconductors C, Si and Ge,
and compound semiconductors GaAs and AlAs.  Furthermore, we study two wide gap insulators
LiF and Ar. Except for Ge, the test set is chosen to match the available many body perturbation theory data. \cite{ShamSchluter, PhysRevB.36.6497, gruning} 
We evaluate both the
direct bandgaps in high symmetry points of the Brillouin zone
and the fundamental indirect band gap where relevant.
In each case the two contributions to the quasiparticle gap, the KS band gap and the discontinuity
$\Delta_{\rm xc}$ are given.
We compare our data to the experimentally observed and
to other calculated results, where available.

In section \ref{sec:discontinuity} the basic concepts are defined
and the Sham--Schl\"uter equation is briefly introduced.
The GLLB model potential is introduced in section
\ref{sec:GLLB} and extended to suit better for
solids and band gap materials, in particular.
In section \ref{sec:gllbdiscontinuity} 
the discontinuity of GLLB potential is discussed. 
Section  \ref{sec:implementation} gives some details about implementation to
the GPAW code.
Finally, the results are given in section \ref{sec:GLLBResults} and conclusions in section \ref{sec:conclusions}.

\section{Quasiparticle band gap}
\label{sec:discontinuity}

\ffbox{Esitellään derivaatan epäjatkuvuuteen liittyvä teoria mahdollisimman
lyhyesti ja ytimekkäästi. Kerrotaan miten tämä johtaa potentiaalin kasvamiseen
kiteissä, kun sinne lisätään elektroni. Myös tarkemmin 
asiaa eri potentiaalien derivaatan epäjatkuvuuksista (OEP-EXX, EXX-RPA).
Ehkä jopa kokonaan tänne kaikki sellaiset introductionista?}

The KS-DFT exchange--correlation (XC) potential of an $N$-electron system
is the functional derivative of the XC energy as
\begin{equation}
v_{\rm xc}(\rr; N) = \left. \frac{\delta E_{\rm xc}[n]}{\delta n(\rr)}\right|_N.
\label{xcpot}
\end{equation}
\noindent
It is continuous with respect to the fractional number of electrons, but
at integer occupations $J$ a discontinuity may emerge as
\begin{equation}
\Delta_{\rm xc} = \Delta_{\rm xc}(\rr) = v_{\rm xc}(\rr; J+\delta) - v_{\rm xc}(\rr; J-\delta),
\label{dxc}
\end{equation}
\noindent
where the limits $\delta \rightarrow 0$ are implied. 
The discontinuity $\Delta_{\rm xc}$ is a constant function of ${\bf r}$.\cite{KSEnergies}

Within the exact DFT, the quasiparticle band gap of an N-electron system,
the difference of the ionization potential (I) and electron affinity (A),
consists of two contributions \cite{KSEnergies, ShamSchluter}
\begin{align}
E^{\rm QP}_{\rm g} & = I - A = E[n_{N-1}]  - 2E[n_N] + E[n_{N+1}] \nonumber \\
& = E^{\rm KS}_{\rm g} + \Delta_{\rm xc},
\end{align}
\noindent
where the first term $E^{\rm KS}_{\rm g} = \varepsilon_{N+1}-\varepsilon_{N}$ is the KS band gap
and the second term is the derivative discontinuity.

First estimates for the derivative discontinuity on real material 
was given by Godby \etal,\cite{PhysRevLett.56.2415} who solved $v_{\rm xc}$ from 
the Sham--Schl\"uter equation\cite{ShamSchluter} 
\begin{eqnarray}
0 = \int d\omega \int d2 \int d3
G_{\rm KS}({\bf r}_1,2;\omega) \nonumber \\
\left\{ \Sigma^{\rm xc}(2,3;\omega) - v_{\rm xc}^{\rm KS}({\bf r}_2)\delta(2-3)
\right\} G(3,{\bf r}_1;\omega),
\label{shams}
\end{eqnarray}
\noindent
by linearization: the interacting Green's function $G$ and $G_{\rm KS}$ were both replaced by the $G_{\rm LDA}$.
The resulting potential is expected to be close to true KS-DFT, thus, leading to
the band gaps equally close. Therefore, we compare GLLB and \GLLBSC~band gaps
to those obtained by Godby \etal \cite{PhysRevLett.56.2415, PhysRevB.36.6497} for C, Si, 
GaAs and AlAs. Later, Gr\"uning \etal\cite{gruning} evaluated using similar methods
for Si, LiF and Ar. We refer to all these data as \trueks ~values, later on.

\section{GLLB exchange and Coulomb correlation}
\label{sec:GLLB}

The exchange and correlation energy functional can be written 
in terms of coupling constant averaged pair correlation function
$\bar{g}_{\rm xc}$ \cite{GLB,GLLB}
\begin{align}
E_{\rm xc}[n] = & \frac{1}{2} \int d \rr_1 \int d \rr_2 n(\rr_1) n(\rr_2) \\  \nonumber 
& \times  v(\rr_1, \rr_2) (\bar{g}_{\rm xc}[n](\rr_1, \rr_2)-1), \\  \nonumber
\label{Exc4}
\end{align}
\noindent
which leads to the exchange--correlation potential in Eq.~\Ref{xcpot}
as \cite{GLB,GLLB}
\begin{equation}
v_{\rm xc}(\rr) = v_{\rm scr}(\rr) + v_{\rm resp}(\rr), \\
 \end{equation}
where the two contributions are
\begin{equation}
v_{\rm scr}(\rr_1) = \int d \rr_2 n(\rr_2) v(\rr_1, \rr_2) (\bar{g}_{\rm xc}[n] (\rr_1,\rr_2)-1) \\
 \end{equation}
and
\begin{align}
v_{\rm resp}(\rr_1) = & \frac{1}{2} \int d \rr_2 \int d \rr_3 n(\rr_2) n(\rr_3) \nonumber \\
& \times v(\rr_2, \rr_3) \frac{\delta \bar{g}_{\rm xc}[n](\rr_2, \rr_3)}{\delta n(\rr_1)}. \nonumber \\
\end{align}

The screening part $v_{\rm scr}(\rr)$ is the Coulombic potential of the XC hole,
corresponding to the Slater potential in the exchange-only case.
Thus, it has a smooth and attractive form.
The response part $v_{\rm resp}(\rr)$ arises from the
pair correlation function response to the density variations.
It is repulsive and short-ranged.
Next, these two parts will be approximated with the help of a GGA functional.

In the original GLLB approach\cite{GLLB} the B88 exchange functional was used, 
because of the correct asymptotic behavior 
($-1/r$) and a parameter fit to atoms. \cite{Becke88}  Obviously, these are important
features for small finite systems.
We choose a modification of PBE functional\cite{PhysRevLett.77.3865} for solids, 
\cite{perdew:136406} PBEsol, instead. It is the ``state of the art`` density-functional, 
to restore the response properties of local-density approximation and the jellium surface energy.
As we deal with the electronic structures of solids, the choice is natural.

In the further work \cite{GLB}, the GLLB screening was completed with a correlation
contribution from the energy density of Perdew and Wang.\cite{PhysRevB.44.13298}
In this work we 
write for the screening potential approximation

\begin{equation}
v_{\rm scr}(\rr) = 2 \epsilon_{\rm xc}^{\rm (PBEsol)}(\rr),
\end{equation}
\noindent
where $ \epsilon^{\rm (PBEsol)}_{\rm xc}$ is the XC energy density.

The exchange response part is the central issue here, and therefore, it deserves a closer look.  First,
within the KLI approximation\cite{KLI} the exchange response potential is written as
\begin{equation}
v_{\rm resp}(\rr) = \sum_{i}^{\rm occ} w_i \frac{|\psi_i(\rr)|^2}{n(\rr)},
\label{respot}
\end{equation}
\noindent
where the coefficients $w_i$ are chosen
self-consistently as
\begin{equation}
w_i = \left< i \right| v_{\rm x}(\rr) - \widehat V^{(HF)}_{\rm x}  \left| i \right>,
\label{elements}
\end{equation}
\noindent
where $\widehat V^{(HF)}_{\rm x}$ is the computationally heavy Fock-operator.

The corresponding approximate exchange response part of GLLB was formulated by Gritsenko \etal \cite{GLLB}
using several physical arguments:
exchange scaling relation, asymptotic behavior and fit to the homogeneous electron gas.
This was carried out by formulation of a simple expression for the orbital dependent function
$w_i$, Eq.~(\ref{elements}), which only depends on KS eigenvalues.

Shift of the external potential by a constant should not have any physical effect, and thus, the function should depend
on the differences of the eigenvalues, only. 
Therefore, the highest occupied eigenvalue $\varepsilon_H$ is taken as a reference $\varepsilon_r$ and we choose
\begin{equation}
w_i = f(\varepsilon_r - \varepsilon_i),
\label{dep}
\end{equation}
with the condition that $f(0) = 0$, as $w_H$ should vanish.\cite{KLI}

Furthermore, the exchange potential has the following scaling property
\begin{equation}
v_{\rm x}[n_\lambda](\rr) = \lambda v_{\rm x}[n](\lambda \rr),
\end{equation}
\noindent
where $n_\lambda = \lambda^3 n(\lambda \rr)$,
while the eigenvalues scale as
\begin{equation}
\varepsilon_i[n_\lambda] = \lambda^2 \varepsilon_i[n(\rr)].
\end{equation}
\noindent
These imply that the function $f$ should scale as
\begin{equation}
f(\lambda^2(\varepsilon_r - \varepsilon_i)) = \lambda f(\varepsilon_r-\varepsilon_i),
\end{equation}
\noindent
which is satisfied by the form
\begin{equation}
w_i = K_{\rm x} \sqrt{\varepsilon_r - \varepsilon_i}.
\label{wi}
\end{equation}

The response potential of the homogeneous electron gas (HEG) is known and it is
\begin{equation}
v_{\rm resp} = \frac{\rm k_F}{2\pi},
\label{heg1}
\end{equation}
\noindent
where the Fermi wave vector is $k_F = (3 \pi^2 n)^{1/3}$.
The corresponding response potential in this approach is
\begin{equation}
v_{\rm resp}^{\rm HEG} =
\frac{V}{8\pi^3} \int_{|{\bf k}|<{\rm k_F}} d{\bf k} \, K_x \sqrt{\varepsilon_r - \varepsilon_{\bf k}},
\label{heg2}
\end{equation}
\noindent
where the difference $\varepsilon_r-\varepsilon_k$ for the electron gas can be written as 
\begin{equation}
\varepsilon_r - \varepsilon_k = (k_F^2/2 + v_{\rm KS}) - (k^2/2 + v_{\rm KS}).
\end{equation}

Setting the right hand sides of \Ref{heg1} and \Ref{heg2} equal, evaluation of the integral yields
the electron gas fitted prefactor
\begin{equation}
K_{\rm x} = \frac{8\sqrt{2}}{3\pi^2} \approx 0.382.
\end{equation}

Gritsenko \etal \cite{GLB} use this same functional form also for the correlation
contribution in the response part and just fit the relevant prefactor $K_{\rm c}$, accordingly.
We choose to use the GGA, again, and the same PBEsol as before, consistently.
As $v_{\rm c} = v_{\rm c,scr} + v_{\rm c,resp}$, we simply write
\begin{equation}
v_{\rm c,resp}^{\rm PBEsol}(\rr) = v^{\rm PBEsol}_{\rm c}(\rr) - 2 \epsilon_{\rm c}^{\rm PBEsol}(\rr).
\label{vcresp}
\end{equation}

Thus, the total \GLLBSC-potential can be finally written as
\begin{align}
v_{\rm \GLLBSC}(\rr) = \, & 2\epsilon_{\rm xc}^{\rm PBEsol}(\rr) \nonumber \\
& +  \sum_i^{\rm occ} K_{\rm x} \sqrt{\varepsilon_r - \varepsilon_i} \frac{|\psi_i(\rr)|^2}{n(\rr)}
+v^{\rm PBEsol}_{\rm c,resp}(\rr).
\end{align}

In summary, the above formulation is an orbital-dependent robust simplification
of the KLI approximation\cite{KLI} to the EXX-OEP \cite{GorlingDxc} following
the guidelines of GLLB\cite{GLLB, GLB} for the exchange.
For correlation, our formulation adds PBEsol correlation,\cite{perdew:136406}
which is consistent with the exchange screening part.

\section{Discontinuity in GLLB+SC}
\label{sec:gllbdiscontinuity}

In this section, we discuss the discontinuity and its origin in response potential.
For our \GLLBSC~has only exchange discontinuity,
the expression for the discontinuity is identical with that of GLLB.
In both the potential  
is not a direct functional derivative of any XC energy functional,
similarly to KLI\cite{KLI} and LHF\cite{sala:5718} approximations.  
However,
due to the similar orbital-dependence all these potentials exhibit the discontinuity on addition of an electron.
In GLLB exchange response approximation, the discontinuity comes with the coefficients $w_i$ in Eq. \Ref{dep}
from their straightforward dependence on the highest occupied electron state.

The reference energy $\epsilon_r$ for particle number N close to integer occupation J can be written as
\begin{equation}
\epsilon_r = \left\{ \begin{tabular}{cc}
 $\epsilon_J$ & $,N \leq J$ \\
 $\epsilon_{J+1}$ & $,N > J$, \\
\end{tabular} \right.
\end{equation}
for when the occupation exceeds J, what was formerly lowest unoccupied molecular orbital (LUMO) becames now the highest occupied.
For the difference of the above and below limits of $v_{\rm x}({\bf r})$ as $N \rightarrow J$,
i.e.~the discontinuity, one obtains straightforwardly
\begin{equation}
\Delta_{\rm x, resp}({\bf r}) = \sum_{i}^{N} K_x \left( \sqrt{\varepsilon_{N+1} - \varepsilon_i}
 - \sqrt{\varepsilon_{N} - \varepsilon_i} \right) \frac{|\psi_i(\rr)|^2}{n(\rr)}.
\label{ddisc}
\end{equation} 

As the above approximation is not a constant, but depends on the space coordinate,
the wave functions would be effected. 
Therefore, to compare with our approach the first order perturbation theory expression leading
to the constant discontinuity should be evaluated as
\begin{equation}
\Delta_{\rm x, resp} = \left< \Psi_{N+1} \right|\Delta_{\rm x, resp}^{\rm GLLB} \left| \Psi_{N+1} \right>.
\label{calcDeltaxc}
\end{equation}

By analysing the term in Eq. (\ref{ddisc}) for different summation indices more closely, we note that it vanishes for $\epsilon_i \rightarrow -\infty$. 
Thus, the dominant contribution from this expression is from neighbourhood of the fermi energy as one would expect.

In addition, we wish to revise a connection between the Sham--Schluter equation
and several approximations to the response potential such as in KLI or in GLLB.
Using similar arguments as those in derivation of KLI, to simplify the Sham--Schl\"uter equation, Eq.~(\ref{shams}),
after cumbersome algebra Casida found an approximative solution to $v_{\rm xc}({\bf r})$ in
terms of the self-energy\cite{PhysRevA.51.2005}
\begin{eqnarray}
v_{\rm xc}({\bf r})
 = \sum_i^N \frac{{\rm Re}\{\psi_i({\bf r}) 
\widehat \Sigma^{\rm xc}(\epsilon_i) 
\psi_i({\bf r}) \}}{n({\bf r})} \nonumber \\
+  \sum_i^N \frac{\left< \psi_i | v_{\rm xc} 
- \widehat \Sigma^{\rm xc}(\epsilon_i)
 | \psi_i \right>{|\psi_i(\rr)|^2}}{n({\bf r})},
\label{ee}
\end{eqnarray}
\noindent
where the latter term is equivalent of the response part of KLI,
if $\Sigma^{\rm xc} \approx \Sigma^{\rm x} = iG_{\rm DFT}v$, ie.~the
x-only self-energy in OEP-EXX formalism where v is the bare coulomb interaction.
By relating Eqs.~(\ref{elements}) and (\ref{wi}),
the response potential of GLLB, and ours, turns out to be an approximation to 
the matrix element in Eq.~(\ref{ee}) as
\begin{equation}
\left< \Psi_i | v_{\rm x} - \widehat \Sigma^{\rm x}(\epsilon_i) | \Psi_i \right> \approx 
K_x \sqrt{\epsilon_r - \epsilon_i}.
\end{equation}

\section{Implementation}

\label{sec:implementation}

We have implemented the GLLB and \GLLBSC~potentials to the grid based projector augmented wave method code GPAW\cite{GPAW}.
It is a pseudo-potential free approach, which allows more accurate and controlled description of electronic structure than
the the conventional pseudo potential approximations. 
For PAW core electrons the frozen-core approximation is used.

The PAW method \cite{PAW,PAW2} is based on a linear transformation, which connects smooth wave functions 
(represented in coarse cartesian grid in GPAW) to the accurate all-electron functions (represented using partial wave set
within each augmentation sphere in GPAW). The transformation and the resulting 
one-particle equation are
\begin{align}
\widehat T \tilde{\Psi}(\rr) & = \Psi(\rr) \\
\widehat T^\dagger \widehat H \widehat T \tilde{\Psi}(\rr) & = E \widehat T^\dagger \widehat T \tilde{\Psi}(\rr)
\end{align}

Details of the transformation are given elsewhere\cite{GPAW}.
Normal approach for deriving the PAW-potential would to take the derivative of the total energy expression,
but since GLLB or \GLLBSC~have no such expression we form the potential analoguously by hand.
The PAW potential consists of a smooth part, which can be chosen ``in priciple'' freely inside the augmentation sphere.
To obtain sufficiently smooth potential, we choose the expression
\begin{eqnarray*}
\tilde{v}_{\rm \GLLBSC}(\rr) = 2 \epsilon^{\rm PBEsol}_x[\tilde{n}(\rr), |\nabla \tilde{n}(\rr)|^2](\rr)
\\
+ \sum_i^{val.} K_G \sqrt{\epsilon_r - \epsilon_i} \frac{|\tilde{\psi}_i(\rr)|^2}
{\sum_i^{val.}|\tilde{\psi}_i(\rr)|^2} \\
+ v_{\rm c}^{PBEsol}[\tilde{n}(\rr), |\nabla \tilde{n}(\rr)|^2](\rr),
\end{eqnarray*}
\noindent
which is clearly identical to all-electron \GLLBSC-potential outside and smooth inside the augmentation spheres. The GLLB potential
is obtained similarly by replacing $\epsilon^{\rm PBEsol}_x$ by $\epsilon^{\rm B88}_x$ and omitting the correlation potential.

The smooth potential requires augmentation sphere corrections to obtain full-potential description
and we calculate the total PAW-Hamiltonian as

\begin{align}
\tilde{\widehat v}_{\rm xc} = \tilde{v}_{\rm xc}(\rr) + \nonumber \\
\sum_a^{atoms} \sum_{ij} | \tilde{p}^a_i \rangle
\left(
\langle \phi_i^a | v_{\rm xc}^a(\rr) | \phi_j^a \rangle
-
\langle \tilde{\phi}_i^a | \tilde{v}^a_{\rm xc}(\rr) | \tilde{\phi}^a_j \rangle
\right)
\langle \tilde{p}^a_j |,
\label{GLLBPAW}
\end{align}

\noindent
where the spherical corrections are performed on a radial logarithmic grid and the smooth part on a sparse real-space grid.
The $\tilde{\phi}_i$ are the partial wave expansions used to generate pseudo density within augmentation sphere
and $\phi_i$ are corresponding all-electron partial waves. The partial waves $\tilde{\phi}^a_i$ and projectors
$\tilde{p}^a_j$ are chosen bi-orthogonal, thus on infinite basis set limit 
$\sum_i | \tilde{p}^a_i \rangle \langle \phi_i^a |$ and it's conjugate are identity operators within the augmentation sphere.
The quanities $v_{\rm xc}^a$ and $\tilde{v}^a_{\rm xc}$ are the radial all-electron xc-potential and radial smooth xc-potential correspondingly.
They are constructed and integrated in 50 radial slices corresponding to Lebedev points in a unit sphere.

For core states, we use the response potential calculated for single atom. In addition, for calculating the discontinuity, we neglect the
shift caused by the core states. This is justified for core states, for their contribution is small due
to reasons described in section \ref{sec:gllbdiscontinuity}.

Calculation of the potential scales as $O(N^2)$ with a small prefactor due to construction of the
response potential, i.e., like evaluation of the density from KS orbitals. 
Thus, the computation scales similarly as the normal local density functional potentials,
with a slightly larger prefactor arising mostly due to a larger number ($\approx 1.5 \times$)
of SCF iterations. In case of a general model potential, this issue is discussed further in Ref.~\onlinecite{gaiduk:044107}.

\section{Results and discussion}
\label{sec:GLLBResults}

To ensure convergence with respect to numerical parameters in calculations, a real space grid
with about 0.11 \AA \ spacing was used for the wave functions.
We also used $17\times 17\times 17$ k-points in the first Brillouin zone corresponding to
the two atom unit cell to find the conduction band minimum (CBM) state required for calculation of the discontinuity accurately.
Note, that this is more than what is needed for convergence in the self-consistent electronic structure.
Consequently, sufficient numerical accuracy in the obtained KS potential and
$\Delta_{\rm x, resp}^{\rm GLLB}$ is guaranteed.
Using these, the single-point band structure calculations were performed
using k-points in high-symmetry points and directions in the first Brillouin zone.

Smooth and all-electron partial waves and the pseudo projector functions were generated on default values provided with the GPAW code\cite{GPAW}.
The frozen-core approximation was used.
In case of Ga the 3d electrons were included into the frozen core to retain comparability to some earlier
pseudopotential calculations, but relaxation of the 3d electrons was tested and found
to have only a minor effect on the band gap.

\begin{figure} [t]
\includegraphics[width=3.5in]{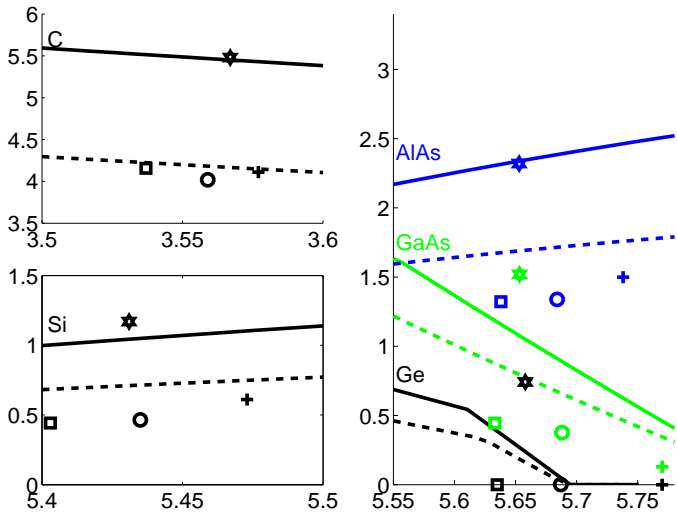}
\caption{(Color online) Fundamental, i.e.~minimum, band gap (in eV) of five semiconductors as a function
of the lattice constant.  The KS gap (dashed curve) and the quasiparticle gaps
(solid curve) from our \GLLBSC~are shown.  The LDA (square), PBE (plus) and
PBEsol (circle) data are shown for their minimum energy lattice constants,
respectively.  Similarly, the experimental data is denoted by the star.
For references, see Table \ref{qpgaps}.}
\label{disps}
\end{figure}

We do not have the total energy functional to minimize in our approach for finding the crystal lattice constants
of our test set semiconductor compounds, consistently.  Therefore, we first consider the evaluated
fundamental (minimum) band gaps
as a function of the lattice constant in the range of LDA, GGA and experimental gaps, shown
in Fig.~\ref{disps}. The possible structural changes due to stress are not taken into account. 
Note, that our primary intention is not to evaluate band gap for materials under stress, but
to aknowledge the fact that the lattice constant has a large effect to the band gap. Therefore the band gap predicted
using relaxed lattice constant depends not only on the xc potential, but also on the energetic properties of xc functional (ie. on the relaxed lattice constant itself).

Also, for comparison the LDA, PBEsol and experimental band gap--lattice constant data is given.
There, the usual tendency of LDA underestimating and GGA overestimating the experimentally
found lattice constants is clearly seen.  The PBEsol is seen to find a lattice constant in between
these two, and in average, closest to the experimental one.
As \GLLBSC~is based on PBEsol, it can be suggested to be used for evaluation of the lattice
constants and other energetics for \GLLBSC, where relevant.

From our \GLLBSC~approach the KS contribution and the total quasiparticle band gaps are shown.
Lattice constant dependence is seen to be weak for C and Si, but stronger for compound semiconductors and Ge.
For GaAs and Ge the lattice constant dependence is strongest and match with experimental gaps is less good.
The gap opens strongly with decreasing lattice constant.
The success with Ge should be noticed, in particular, as the LDA and GGA
do not open the gap, at all.  The other cases show a good match with the experimental
band gap in a large range of lattice constants.

\begin{figure} [t]
\includegraphics[width=3.5in]{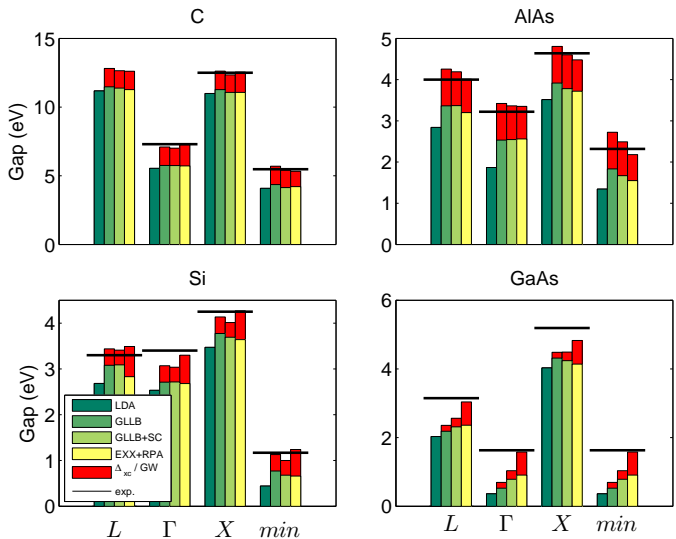}
\caption{(Color online) Kohn--Sham band gap (three leftmost, {\color{green}green}) and
the discontinuity (top, {\color{red}red}) contribution
to the quasiparticle (total) gap from our LDA, GLLB exchange-only and \GLLBSC~calculations.
The ``True DFT'' KS gaps (rightmost, {\color{yellow}yellow}) with the discontinuity 
from GW (top, {\color{red}red}) are shown for comparison.}
\label{bargaps}
\end{figure}

\begin{table} 
\begin{tabular}{cccccc}
\hline
Compound & LDA & GLLB$^a$ & \GLLBSC$^a$ & KS/GW$^b$ & exp.$^c$ \\
\hline
  C & 4.09 & 4.36/5.70 & 4.14/5.41 & 4.21/5.33 & 5.48 \\
  Si & 0.44 & 0.77/1.13 & 0.68/1.00 & 0.66/1.24 & 1.17 \\
  Ge & 0.00 & 0.00/0.00 & 0.21/0.27 & NA & 0.74 \\
  AlAs & 1.34 & 1.83/2.72 & 1.67/2.49 & 1.55/2.18 & 2.32$^d$ \\
\hline
  GaAs & 0.36 & 0.53/0.69 & 0.79/1.04 & 0.91/1.58 & 1.63$^d$ \\
  LiF & 8.78 & 11.20/15.38 & 10.87/14.96 & 9.3/13.5 & 14.2 \\
  Ar & 8.18 & 9.9/14.46 & 10.3/14.97 & 8.8/13.1 & 14.2 \\
\hline
\hline
\end{tabular}
\caption{$^a$ The minimum KS band gaps/the fundamental band gap with discontinuity from GLLB exchange-only and \GLLBSC~calculations using the experimental lattice constants given in the text.
$^b$ The KS-band gap based on Sham--Schl\"uter GW self-energy/GW 
quasiparticle band gap\cite{PhysRevB.36.6497, gruning}. 
$^c$ Experimental values for C, Si, AlAs, GaAs from Ref. 8 
and references there in. For Ge we used 0K value from Ref. 32. 
LiF and Ar values from Ref. 7. 
$^d$ effect of spin-orbit splitting removed (see Ref. 8
~for details). All units in eV.}

\label{qpgaps}
\end{table}

From now on we restrict our analysis and discussion to the calculated band gaps using
the following experimental lattice constants: C(3.567)\cite{vol1}, Si(5.431)\cite{vol1}, Ge(5.658)\cite{vol1},
GaAs(5.653)\cite{vol1}, AlAs(5.661)\cite{vol2}, LiF(4.024)\cite{LiF} and Ar(5.260)\cite{Ar},
all in \AA ngstroms.

In Table \ref{qpgaps} we list our calculated KS band gaps for LDA, GLLB and \GLLBSC. These
are to be compared with the band gap values obtained with potential from linearized Sham--Schl\"uter equations.
Both GLLB and \GLLBSC~yield KS band gaps close to these values. Comparing \GLLBSC~and GLLB, \GLLBSC~shifts the band gaps to the right
direction with all materials except for Ar. Furhtermore, we note that the obtained band gaps are much closer to expected KS values
than the approach by Tran. et. al due to their choice of the fitting objective\cite{tran:226401}.

Furthermore, In Table \ref{qpgaps} we list the calculated quasiparticle band gaps with added discontinuity using GLLB and \GLLBSC~potentials.
The GLLB and \GLLBSC~use an electron gas based response potential resulting in a discontinuity which gives good quasiparticle band gaps to be compared
with the GW and experimental results. We find this to be a remarkable result considering the fact that the quasiparticle
band gaps evaluated from OEP-EXX potential are disasterously overestimated\cite{GorlingDxc}. 

In Fig.~\ref{bargaps} we extend our analysis to all of the calculated 
direct KS band gaps of our test case semiconductors at the special symmetry points
in the Brillouin zone.  The discontinuities are shown in red. It should be noted that
Kohn--Sham DFT with added discontinuity guarantees only the fundamental band gap to be correct, 
for it is the only quantity which is a ground state property in the band structure.
However, we approximate also other band gaps by adding the calculated discontinuity also to them as shown in Fig.~\ref{bargaps}.
Good match with the experiments is again seen, except
for GaAs, as in Fig.~\ref{disps} for the fundamental band gaps.

\begin{table} [t]
\begin{tabular}{ccccccc}
Comp. & Gap & LDA$^a$ & GLLB$^a$ & \GLLBSC$^a$ & "True"$^b$ & LDA$^b$ \\
\hline
Si & $\Gamma \rightarrow \Gamma $ & 2.53 & 2.71 & 2.72 & 2.6 & 2.6 \\
  & $\Gamma \rightarrow X$ & 0.58 & 0.91 & 0.81 & 0.6 & 0.7 \\
  & $\Gamma \rightarrow L$ & 1.47 & 1.88 & 1.88 & 1.5 & 1.5 \\
\hline
LiF & $\Gamma \rightarrow \Gamma $ & 8.78 & 11.2 & 10.9 & 9.3 & 8.9 \\
  & $\Gamma \rightarrow X$ & 14.4 & 17.1 & 16.8 & 15.3 & 14.8 \\
  & $\Gamma \rightarrow L$ & 10.3 & 13.4 & 13.1 & 11.1 & 10.6 \\
\hline
Ar & $\Gamma \rightarrow \Gamma $ & 8.18 & 9.9 & 10.3 & 8.8 & 8.2 \\
  & $\Gamma \rightarrow X$ & 10.9 & 12.3 & 12.7 & 11.4 & 10.6 \\
  & $\Gamma \rightarrow L$ & 11.1 & 12.5 & 12.8 & 11.5 & 11.0 \\
\hline
\hline
\end{tabular}
\caption{Kohn--Sham band gaps of high symmetry points with respect to $\Gamma$-point for Si, LiF and Ar.
$^a$ This work calculated using GPAW code\cite{GPAW}.
$^b$ EXX-RPA and LDA gaps calculated by Gr\"uning \etal \cite{gruning}. All units in eV.}
\label{silifar}
\end{table}

\begin{figure} [b]
\includegraphics[width=3.5in]{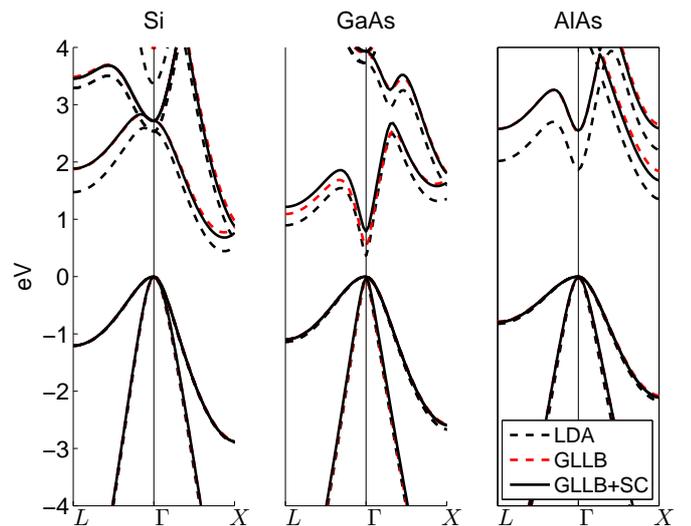}
\caption{(Color online) Calculated Kohn--Sham band structures (without discontinuity)
of compounds Si, GaAs and AlAs using LDA, GLLB and \GLLBSC~approaches.}
\label{cfig}
\end{figure}

\begin{figure} [b]
\includegraphics[width=3.5in]{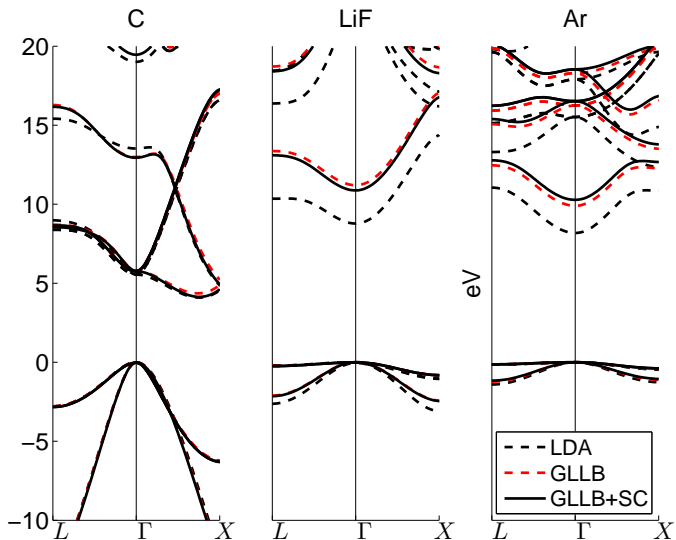}
\caption{(Color online) Calculated Kohn--Sham band structures  (without discontinuity)
of C, LiF and Ar using LDA, GLLB and \GLLBSC~approaches.}
\label{lifig}
\end{figure}

Next, we analyze the Kohn-Sham contribution to the band structure.
First in Table \ref{silifar}, we compare the KS band gaps of highest symmetry points of the first Brillouin zone for Si, LiF and Ar crystals.
LDA gives systematically the lowest gaps underestimating the true KS gaps
while GLLB exchange-only or \GLLBSC~make a slight overestimation.  However, the variation is small, and thus,
not essential. The unaccuracy for comparing our projector augmented wave approach and the pseudo potential approach used for KS gaps
is probably larger than the differences in band gaps.

Finally, we evaluate the Kohn--Sham band structures of the test compounds to
analyze also the dispersion around valence and conduction bands.
In Figs.~\ref{cfig} and \ref{lifig} we compare LDA, GLLB and \GLLBSC~approaches.
The constant discontinuity is removed for clarity.

Overlap of the bands from these approaches is close to perfect in the valence bands and below.
There are no significant differences in the dispersion at the CBM or above, either.
The small differences in the KS gap, see Table \ref{silifar}, make a just rigid shift
of the bands, only.  This behavior seems to be similar in all considered cases.

We argue, that the full potential of model potentials is currently not used based on our positive experience
for simple KS eigenvalue dependent GLLB exchange response potential for predicting the derivative
discontinuity of the exchange-correlation energy functional. 
There are various local and global quantities which are fast to evaluate and 
could be used to construct a mapping between them and electron gas based expressions for potential.
These include quanities such as the eigenvalues (or some other expectation values), wave functions and their gradients
etc. The most simplest approach would be to make the response potential exact at electron gas limit
by fitting the function f in Eq. \ref{dep} to $v_{\rm c,resp}$ for electron gas.
For a first hint, we releated the GLLB response potential to the Casida's approximation to the Sham--Schl\"uter equation.

\section{Conclusions}
\label{sec:conclusions}

We have demonstrated how the derivative discontinuity at the integer occupation numbers
can be included into a simple semilocal orbital-dependent exchange--correlation potential.
Our approach, \GLLBSC, is based on the
GLLB type exchange of Gritsenko \etal \cite{GLLB} and 
PBEsol correlation \cite{perdew:136406}, where the former is responsible for bringing in
the discontinuity in its "response part".  

We have analyzed the roles of the two parts to the evaluated total quasiparticle band gap: Kohn--Sham gap
and the discontinuity contribution. Both GLLB and \GLLBSC~potentials contain only dicontinuous exchange potential,
but nevertheless the agreement with experimental results is remarkable compared to computationally more expensive EXX
approach, where the quasiparticle band gap is essentially same than the Hartree-Fock band gap.

The evaluated fundamental band gaps for our test set, typical semiconductors and dielectrics,
match surprisingly well to the experimental data and to those from more sophisticated approaches.
However, the computational efforts needed for \GLLBSC~are about the same as for a typical GGA calculation, only.

In short, we have demonstrated a computational approach to solve the "band gap problem"
of semiconductors and shown that it gives close to correct band gaps.

\section{Acknowledgements}
We thank the Academy of Finland (MODEX project), the National Graduate School
of Materials Physics for financial support and CSC for computational resources.
\bibliography{sources}

\end{document}